\documentclass[aps,pra,twocolumn,amsmath,amssymb,nofootinbib,superscriptaddress]{revtex4-2}
\usepackage{todonotes}
\usepackage{times}
\usepackage{graphics}
\usepackage{dcolumn}
\usepackage{bm}
\usepackage{float}
\usepackage{amsmath}
\usepackage{amsthm}
\usepackage{braket}
\usepackage{indentfirst}
\usepackage{xr}
\usepackage[colorlinks]{hyperref}
\usepackage{xcolor}
\usepackage{paralist}
\usepackage{tabularx}
\usepackage{multirow}
\usepackage{booktabs}
\usepackage{mathtools}   

\DeclarePairedDelimiter{\floor}{\lfloor}{\rfloor}
\usepackage[normalem]{ulem} 
\usepackage{array}
\usepackage{longtable}
\usepackage{ragged2e}

\renewcommand\vec{\mathbf}
\usepackage{subfigure}
\usepackage[ruled,vlined,linesnumbered,algo2e]{algorithm2e}
\usepackage{algpseudocode}
\usepackage{amsmath}
\usepackage{amssymb}
\usepackage{verbatim}
\usepackage{makecell}
\usepackage[normalem]{ulem}

\begin{document}
\title{Suppressing Measurement Noise in Logical Qubits Through Measurement Scheduling}
\author{Xiao-Yue Xu} 
\affiliation{Henan Key Laboratory of Quantum Information and Cryptography, Zhengzhou, Henan 450000, China}

\author{Chen Ding}
\affiliation{Henan Key Laboratory of Quantum Information and Cryptography, Zhengzhou, Henan 450000, China}


\author{Wan-Su Bao}
\email{bws@qiclab.cn}
\affiliation{Henan Key Laboratory of Quantum Information and Cryptography, Zhengzhou, Henan 450000, China}

\date{\today}
\begin{abstract}
  Quantum error correction is essential for reliable quantum computation, where surface codes demonstrate high fault-tolerant thresholds and hardware efficiency. However, noise in single-shot measurements limits logical readout fidelity, forming a critical bottleneck for fault-tolerant quantum computation. We propose a dynamic measurement scheduling protocol that suppresses logical readout errors by adaptively redistributing measurement tasks from error-prone qubits to stable nodes. Using shallow entangled circuits, the protocol balances gate errors and measurement noise. This is achieved by dynamically prioritizing resource allocation based on topological criticality and error metrics. When addressing realistic scenarios where temporal constraints are governed by decoherence limits and error-correction requirements, we implement reinforcement learning (RL) to achieve adaptive measurement scheduling. Numerical simulations show that logical error rates can be reduced by up to 34\% across code distances for 3 to 11, with enhanced robustness in measurement-noise-dominated systems. Our protocol offers a versatile, hardware-efficient solution for high-fidelity quantum error correction, advancing large-scale quantum computing.
\end{abstract}
\maketitle

\section{Introduction}
Quantum error correction (QEC)~\cite{Cory1998Experimental,Knill2000Theory,Chiaverini2004Realization,Reichardt2024Logical,Paetznick2024Demonstration} is a critical framework for protecting quantum information from decoherence and operational errors, enabling reliable quantum computation. Among various quantum error-correcting (QEC) codes, the surface code has emerged as a particularly promising candidate, demonstrating superior fault-tolerant thresholds and unique compatibility with 2D lattice architectures requiring only nearest-neighbor qubit interactions~\cite{Google2024Quantum,Acharya2023Suppressing,Fowler2012Surface,Zhao2022Realization,Fowler2010Surface,Horsman2012surgery}.
Its fault tolerance stems from topological protection against local errors, achieving a high error threshold ($\sim$1\%) while requiring only nearest-neighbor parity measurements. The code distance $d$ exhibits exponential suppression of logical error rate $p_L$ (scaling as $p_L \propto (p/p_\text{th})^{d/2}$) when physical error rate $p$ remains below the threshold $p_\text{th}$. This combination of hardware-efficient geometry, robust error suppression, and low operational overhead makes it a cornerstone for scalable quantum computation.

\begin{figure}[t]
  \centering
  {\includegraphics[width=0.35\textwidth]{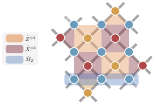}} 
  \caption{{\textbf{A rotated surface code lattice and stabilizer measurement circuits.} The two-dimensional array of the rotated surface code which represents a logical qubit with code distance $d=3$. Data qubits (blue) store quantum information, while measurement qubits (orange/red) execute stabilizer operations—orange for $Z$-type stabilizers $\hat{Z}^{\otimes 4}$ and red for $X$-type stabilizers $\hat{X}^{\otimes 4}$, with corresponding colored blocks highlighting qubit subsets involved in each stabilizer circuit. Boundary conditions are implemented via blue blocks enclosing data qubits at the code's smooth boundary, where direct $Z$-basis physical measurements resolve the logical qubit state.}}
  \label{fig1}
\end{figure}

The surface code operates on a lattice with alternating {\textit{smooth}} ($Z$-type) and {\textit{rough}} ($X$-type) boundaries. Two types of stabilizers, $X$-type and $Z$-type, are measured periodically to detect errors via syndrome extraction, as illustrated in Fig.~\ref{fig1}.  
The surface code’s fault tolerance hinges on two complementary pillars: robust error correction based on stabilizer measurement and precise logical state readout. In planar code architectures~\cite{Horsman2012surgery}, the logical $Z$ are evaluated by parity through horizontal data-qubit paths traversing smooth boundaries, requiring high-fidelity readout of boundary-adjacent data qubits. When considering the measurement error during these two process, stabilizer measurements effectively suppress single-shot readout errors by leveraging temporal redundancy, typically through 5–15 rounds of syndrome averaging. This redundancy allows the quantum state to be preserved with high fidelity over extended periods.  However, the final readout for logical state measurement is irreversible and collapses the quantum state, permanently forfeiting the opportunity for multi-round error mitigation. To harness the surface code’s full potential, noise-resilient logical readout protocols must bridge this gap, reconciling the destructive nature of direct measurement with the stringent fidelity demands of scalable quantum computation.
\begin{figure*}[htbp!]
  \centering
  {\includegraphics[width=1\textwidth]{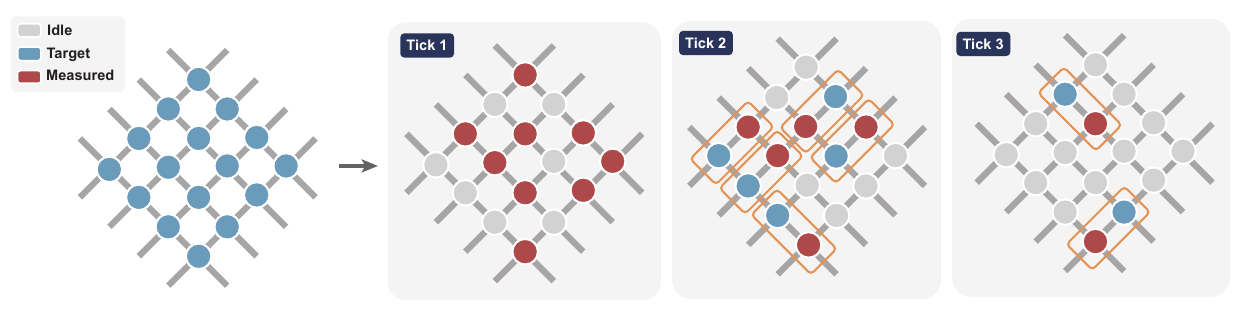}} 
  \caption{{\textbf{The illustration of measurement scheduling protocol in a 2D chain quantum topological structure.} The left panel depicts a 2D chain-type quantum topological structure, where blue marks indicate qubits requiring measurement. Arrows outline the scheduling process across three sequential time steps (ticks). Gray circles represent qubits idle during their respective tick, while blue circles denote measurement targets in the current tick. The orange box demonstrates the measurement transfer modality: a shallow quantum circuit reroutes measurement tasks (blue) to adjacent qubits (red markers within the box). After three ticks, all qubits are measured.}}
  \label{fig2}
\end{figure*} 
In this work, we propose a measurement scheduling (MS) protocol that directly targets the suppression of logical readout errors by redistributing measurement tasks across subsystems through measurement transfer modalities, as illustrated in Fig.~\ref{fig2}. This approach is motivated by the hardware asymmetry in modern quantum architectures. 
While two-qubit gate fidelities are compatible with fault-tolerant thresholds~\cite{Foxen20202020Demonstrating,Yuan2020High,V2021High,Sung2021Realization,Walter2017Rapid}, single-shot readout errors remain at significantly higher levels with up to an order of magnitude discrepancy, constituting a critical performance limitation~\cite{Bultink2020Protecting,Linke2017FaultFault,Chen2023Transmon}. Moreover, unlike gate errors that can be mitigated through techniques such as dynamical decoupling, effective suppression of readout-induced errors remains particularly challenging~\cite{Acharya2023Suppressing,Pino2021Demonstration,Tantivasadakarn2023Hierarchy,Smith2023Scaling}.  

By reframing the quantum readout task as a resource allocation problem, our MS protocol reroutes measurement tasks from error-prone qubits to stable neighboring nodes. This protocol resolves qubit allocation conflicts through dynamic prioritization, guided by real-time error metrics and topological criticality. We first introduce an easily implementable local MS scheme that does not account for the waiting times linked to qubit allocation conflicts. When considering temporal constraints such as error correction deadlines and qubit decoherence limits, the protocol integrates reinforcement learning (RL)~\cite{Mnih2015Human,Silver2016Mastering,Silver2017Mastering,Sutton1999Policy,John2015Trust,schulman2017proximal} to iteratively refine the allocation. This temporal-aware protocol achieves adaptive measurement scheduling while preserving conflict resolution integrity, inherently balancing computational tractability and scheduling precision. We perform extensive numerical experiments exploring the performance landscape of the proposed MS framework, validating both the practical efficacy and theoretical advantages of our approach.

\section{Measurement Transfer Modalities}\label{sec2}
We first introduce measurement transfer modalities as the schedulable units of our MS protocol, designed to work seamlessly with fault-tolerant quantum architectures. These modalities shift measurement tasks from error-prone qubits to high-fidelity subsystems using shallow entangled circuits, reducing overall readout errors on quantum processors with higher gate fidelity. 

\subsection{General measurement transfer modalities}
Within our MS framework, these modalities are divided into two basic types of logical readout operations: multi-qubit parity measurement and single-qubit measurement. 
The parity measurement of $n$ qubits, which assesses whether the number of the number of $|1\rangle$ states in their measurement outcomes is even or odd, can be encoded in the expectation value of the $\hat{Z}^{\otimes n}$ operator. Here, eigenvalues $+1$ and $-1$ of $\hat{Z}^{\otimes n}$ correspond to even and odd parity, respectively.  
We define the single-shot measurement operator as $\mathcal{M}[\hat{O}](\cdot) \in {0,1}$, representing the single-qubit outcome of observable $O$ on its computational basis. Then for $n$-qubit observable $\hat{Z}^{\otimes n}$, $\mathcal{M}[\hat{Z}^{\otimes n}] $ returns the parity bit $p_{\vec{b}}=\left(\sum\limits_{i=1}^{n} b_i\right) \mod{2}$, where $|\vec{b}\rangle=|b_1b_2,...,b_n\rangle$ ($b_i \in \{0,1\},i=1,2,...,n$) is the collapsed state after measurement. 
We then propose the multi-qubit parity measurement $\mathcal{P}[\cdot]$ on qubit $Q_1$ to $Q_n$, which employs the entangled circuit $U_{2^n-1}$:
\begin{equation}\label{eq-U}
  U_{2^n-1}=\prod_{i=n}^1\text{CNOT}_{i+1,i},
\end{equation} 
to sample the $n$-qubit parity by sampling the compressed state $\sigma_{2^n-1}$ in the $Z$-basis:
\begin{equation}
  \begin{aligned}
    \mathcal{P}[Q_{1:n}\to Q_1] &\coloneqq \mathcal{M}[\hat{Z}_1]\cdot U_{2^n-1}[Q_{1:n} \to Q_1]   \\
    &\equiv \mathcal{M}[\hat{Z}^{\otimes n}],
  \end{aligned}
\end{equation}
where we assume $Q_1$ to be the measured qubit without loss of generality.  

Unlike parity measurement, which can bypass measurement noise for $n-1$ qubits, single-qubit measurement reduce overall readout error by strategically redirecting measurement to low-error qubit based on measurement error profiling. 
For a qubit pair $(Q_1,Q_2)$ with target qubit $Q_2$, given asymmetric readout errors with better $Q_1$, we propose the single-qubit measurement $\mathcal{L}[\cdot]$:
\begin{equation}
  \begin{aligned}
   \mathcal{L}[Q_2\to Q_1]&\coloneqq \mathcal{P}[Q_{1,2}\to Q_1]\cdot \mathcal{MR}[Q_1]\\
   &\equiv \mathcal{M}[\hat{Z}_2]\cdot\mathcal{M}[\hat{Z}_1],
  \end{aligned}
\end{equation}
where the measure-reset operator $\mathcal{MR}[\cdot]$ on $Q$ of multi-qubit state $\rho$ is defined as 
\begin{equation}
  \mathcal{MR}[Q] : \rho \mapsto \left( b_Q, |0\rangle\langle 0|_Q \otimes \text{Tr}_Q[\rho] \right)
\end{equation}
with the measurement outcome $b_Q \in \{0,1\}$ and Tr$_p$ referring to partial trace over $Q$. And when $Q_1$ is an auxiliary qubit initialized to $|0\rangle$, the operator $\mathcal{MR}[Q_1]$ can be directly omitted.  
The single-qubit measurement can be easily extended to $n$-qubit measurement, with the detailed algorithm and proof presented in Supplemental materials.

These two basic algorithms can be adapted from $Q_1$-redirected transfer to subsystems, enabling systematic error mitigation and meeting diverse readout requirements through dynamic resource allocation.
 
\begin{figure*}[htbp!]
  \centering
  {\includegraphics[width=0.82\textwidth]{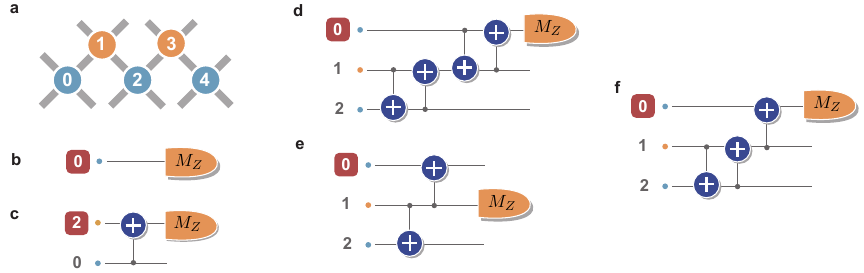}} 
  \caption{{\textbf{Topological qubit connectivity in surface code lattice with circuit designs for measurement transfer modalities.} \textbf{a} Light blue and orange circles denote data and measurement qubits, respectively, arranged in alternating rows. Light grey interconnects depict permitted CNOT gates between non-interconnectable data-measurement qubit pairs lacking intratype connectivity. \textbf{b}-\textbf{f} Circuit implementations for five measurement modalities: $\mathrm{D\text{-}M},\mathrm{MR\text{-}M},\mathrm{DR\text{-}M}, \mathrm{MR\text{-}PM}, \mathrm{DR\text{-}PM}$. The qubits enclosed by the red square denote the target qubits, and those subjected to the orange $M_Z$ measurement operation are the qubits to be measured.}}
  \label{fig3}
\end{figure*}

\subsection{Surface code-specialized modalities} 
We now delve into the specific implementation of measurement transfer modalities for logical qubit readout within the surface code framework, which imposes critical constraints on data qubit measurement to ensure both logical state readout and syndrome extraction.  
We begin with formalize the readout task for a logical qubit.
Let $\mathcal{L}=(D,M,E)$ represent the surface code lattice, 
\begin{itemize}
  \item $D=\{d_{i,j}\}$: Data qubits arranged in the 2D grid (row $i$, column $j$).
  \item $M=\{m_{i,j}^Z,m_{i,j}^X\}$: Measurement qubits ($Z$- and $X$-type stabilizers) interleaved with data qubits.
  \item $E$: Edges defining allowed CNOT gate connections (data-measurement qubit pairs with no intra-data or intra-measurement connectivity.),
\end{itemize} 
as illustrated in Fig.~\ref{fig3} \textbf{a}.
The logical readout of a surface code involves two intertwined objectives:
\begin{itemize}
  \item \textbf{Logical $Z$ measurement}: the smooth boundary (e.g., bottom row) data qubits $\{d_{2,j}\}$ require collective parity to resolving the outcome of $\hat{M}_Z$.
  \item \textbf{$X$-type stabilizer syndromes}: each 
  $X$-type stabilizer (centered on $m_{i,j}^X$) requires the parity of $m_{i,j}^X$ and its connected data qubits (four in the bulk, two on boundaries). Bulk data qubits, participating in multiple stabilizers, need individual measurement for computing all required parities.
\end{itemize} 
The dual requirement creates a conflict: boundary measurements need coordinated parity extraction, while bulk measurements demand precise individual readouts. These competing demands limit measurement grouping due to the spatial distribution of data qubits, arising from two key observations:
\begin{itemize}
  \item \textbf{Boundary Groupability}: on the bottom row (excluding the last qubit), pairs $\{(d_{1,2i},d_{1,2i+1})\}_i$ with $i=0,1,...\floor{\frac{d}{2}}$ share membership in a single $X$-type stabilizer. Similarly, on the top row, pairs $\{(d_{2d,2i+1},d_{2d,2i+2})\}_i$ with $i=0,1,...\floor{\frac{d}{2}}$ can be grouped.
  \item \textbf{Bulk Ungroupability}: all bulk data qubits and non-paired boundary qubits participate in multiple stabilizers. This overlapping membership of different stabilizers prohibits grouping. 
\end{itemize}

Based upon these observations, we define the groupable set $\mathcal{G}$ which allows parity measurement:
\begin{equation}
  \mathcal{G}=\{(d_{1,2i},d_{1,2i+1})\}_{i=0}^{\floor{\frac{d}{2}}} \cup \{(d_{2d,2i+1},d_{2d,2i+2})\}_{i=0}^{\floor{\frac{d}{2}}}.
\end{equation}
And all qubits outside $\mathcal{G}$ require single-qubit measurement.

Then we can establish the surface code-specialized set of modalities:
\begin{equation}
  \mathcal{S}_\text{modes} = \{\mathrm{D\text{-}M}, \mathrm{DR\text{-}M}, \mathrm{MR\text{-}M}, \mathrm{DR\text{-}PM}, \mathrm{MR\text{-}PM}\},
\end{equation} 
where we define:

\begin{itemize}
  \item \textbf{Direct Measurement (D-M)}: projective $Z$-basis measurement on target qubit $d_{i,j}\in D$.
    \begin{equation}
      \mathrm{D\mbox{-}M}(d_{i,j}) \coloneqq \mathcal{M}[\hat{Z}_{i,j}](d_{i,j}) \in \{0,1\}.
    \end{equation}
 
  \item \textbf{$D$-Redirected Measurement (DR-M)}: single-qubit measurement for target qubit $d_{i,j}\in D$ with the adjacent data qubit $d_{p,q} \in D$ to be measured, and the interleaved measurement qubit $m_{k,l}^{\sigma}\in M$ ($\sigma=X/Z$) where $(d_{i,j}, m_{k,l}^{\sigma}),(d_{p,q}, m_{k,l}^{\sigma}) \in E$ and $d_{p,q},m_{k,l}^{\sigma}$ have been reset in $|0\rangle$ states.
  \begin{equation}
    \mathrm{DR\mbox{-}M}(d_{i,j}) \coloneqq \mathcal{L}[d_{i,j}\to d_{p,q}](d_{i,j}) \in \{0,1\}
  \end{equation}
  
  \item \textbf{$M$-Redirected Measurement (MR-M)}: single-qubit measurement for target qubit qubit $d_{i,j}\in D$ with adjacent idle qubit $m_{k,l}^{\sigma}\in M$ to be measured, where $(d_{i,j}, m_{k,l}^{\sigma}) \in E$ and $m_{k,l}^{\sigma}$ has been reset in $|0\rangle$ state.
    \begin{equation}
      \mathrm{MR\mbox{-}M}(d_{i,j}) \coloneqq \mathcal{L}[d_{i,j}\to m_{k,l}^{\sigma}](d_{i,j}) \in \{0,1\}
    \end{equation} 

  \item \textbf{$D$-Redirected Parity Measurement (DR-PM)}: Parity measurement of adjacent pair $d_{i,j},d_{p,q} \in \mathcal{G}$ with interleaved measurement qubit $m_{k,l}^\sigma$, where $m_{k,l}^\sigma$ has been reset in $|0\rangle$ state. We assume $d_{p,q}$ to be measured.
  \begin{equation}
    \mathrm{PE\mbox{-}D}(d_{{i,j},{p,q}}) \coloneqq  
     \mathcal{P}[d_{{i,j},{p,q}}\to d_{i,j}](d_{{i,j},{p,q}})\in \{0,1\} 
  \end{equation}

  \item \textbf{$M$-Redirected Parity Measurement (MR-PM)}: Similarly, but we assume $m_{k,l}^\sigma$ to be measured.
  \begin{equation}
    \mathrm{PE\mbox{-}M}(d_{{i,j},{p,q}}) \coloneqq  
     \mathcal{P}[d_{{i,j},{p,q}}\to m^{\sigma}_{k,l}](d_{{i,j},{p,q}})\in \{0,1\} 
  \end{equation}
  
\end{itemize}
 
Here, we incorporate the modality $\mathrm{D\text{-}M}$ to ensure completeness, despite its exclusion from the measurement transfer algorithm. In Fig~\ref{fig3} \textbf{b}-\textbf{f}, we illustrate the implementation circuits corresponding to the five measurement modalities specialized for surface code.

\section{Local Measurement scheduling for logical qubit readout}
Building upon the foundational scheduling units of measurement transfer modalities, their assignment through modality indices can be formalized as a constrained integer program. The key decision variables and parameters are as follows:
\begin{itemize}
\item $S_{i,j}^k$ denotes the $k$-th modality in $\mathcal{S}$ for data qubit $d_{i,j}\in D$.
\item $\{x_{i,j}^k\} \in \{0, 1\}$ indicates whether modality $S_{i,j}^k$ is chosen.
\item $R_{i,j}^k$ denotes the ensemble of participating physical qubits for $S_{i,j}^k$.
\item $K_{i,j} \in \mathbb{N}$ is the number of available measurement modalities depending on its lattice position.
\end{itemize}
When temporal constraints are relaxed in MS, conflicting qubit allocations ($\mathcal{R}_{i,j}^k \cap \mathcal{R}_{p,q}^{k'} \neq \varnothing$) 
are resolved through iterative temporal deferral, reducing the integer programming formulation: 
\begin{equation}\label{eq-inpro-formulation}
  \underset{{x_{i,j}^k}}{\text{minimize}} \sum_{i,j} \sum_{k=1}^{K_{ij}} x_{i,j}^k \cdot \mathcal{C}(S_{i,j}^k),
\end{equation}
  \textit{subject to:}
\begin{equation}\label{eq-conflict}
\sum_{k=1}^{K_{ij}} x_{i,j}^k = 1, \quad \forall d_{i,j} \in \mathcal{D} 
\end{equation}
to a local greedy optimization framework where modality selection $S_{i,j}^k$ is performed for each data qubit based on optimal local criteria. 
\begin{figure*}[t]
  \centering
  {\includegraphics[width=0.95\textwidth]{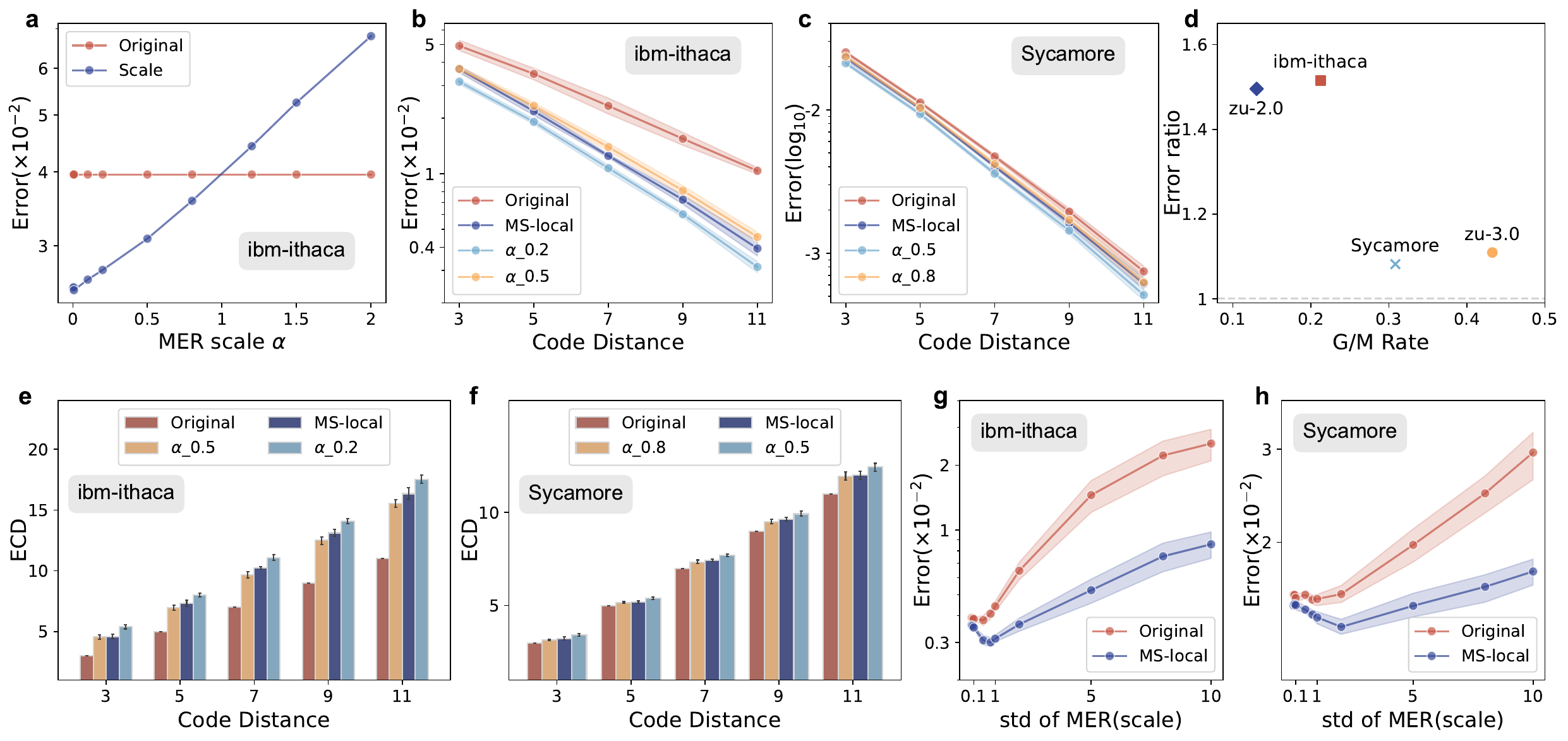}} 
  \caption{{\textbf{The performance of memory experiment for surface code with local measurement scheduling (MS-local).}} In the simulation experiments, we generate random measurement noise distributions based on several actual devices, \textit{ibm-ithaca}, \textit{Sycamore}, \textit{zuchongzhi-2.0} and \textit{zuchongzhi-3.0}. \textbf{a} The logical error rate  versus measurement error rate (MER) scaling for data qubits under fixed measurement-qubit MER and selective data-qubit MER increase. Here, the data-qubit MER scaling parameter $\alpha$ represents synthetic error suppression: $\alpha=1.0$ corresponds to the original hardware MER, while $\alpha=0.5$ simulates a 50\% MER reduction.
  \textbf{b-c} Logical error rate versus code distance from 3 to 11 (with $r=7$) for surface code under: Original direct readout (Original), Original with threshold MER scales bracketing MS-local performance (e.g. $\alpha=0.2$ and $\alpha=0.5$ in \textbf{b}), and MS-local, simulated using error metrics from two typical devices. \textbf{d} The benchmark outcome of the error ratio (Original devided by MS-local) across four quantum devices. The Gate-to-Measurement error ratio (G/M$\equiv\epsilon_m/\epsilon_g$) reflects qubit noise asymmetry---lower G/M ratios correlate with higher error ratios which indicates greater performance gains of MS-local when measurement errors dominate gate errors. The light gray dashed line at ratio 1 represents the threshold where MS-local begins to show its advantage.   
  \textbf{e-f} The effective code distance (ECD) with varying code distance. The ECD of Original equals code distance $d$, while others are calculated via error$=a(p/p_\mathrm{th})^{\text{ECD}/2}$ with parameters $a$ and $p/p_\mathrm{th}$ fitted from Original's error-$d$ data. \textbf{g-h} The logical error rate of MS-local with varying scale of std of MER (equals to true value when scale is 1) across two typical devices. All the data in the figure presents average results with 95\% confidence interval error bars.} 
  \label{fig4}
\end{figure*} 
We then formulate a local measurement scheduling (MS-local) protocol that employs a local greedy approach to iteratively minimize cost through traversal through the lattice structure.

For each data qubit $d_{i,j}\in D$, we enumerate independent modalities $\{S_{i,j}^k\}_k$ based on the cost evaluation using the metric:     
\begin{equation}\label{eq-local cost1}
  \mathcal{C}_\text{ind}(S_{i,j}^k)=\epsilon_m(d_{i,j}^k)+\mathcal{N}(S_{i,j}^k)\cdot \epsilon_g, \ k\in \{1,2,3\},
\end{equation}
where $d_{i,j}^k$ represents the qubit to be measured in $S_{i,j}^k$, $\epsilon_m$ and $\epsilon_g$ denote the measurement error rate (MER) and gate error rate (GER), respectively. And $\mathcal{N}(S_{i,j}^k)$ is the number of two-qubit gates utilized in the entangled circuit for $S_{i,j}^k$ as given in Fig.~\ref{fig3}, which is 
\begin{equation}\label{eq-gate_num1}
  \mathcal{N}(S_{i,j}^k)=
  \left\{
    \begin{aligned}
        &0,\ S_{i,j}^k=\mathrm{D\text{-}M} \\
        &1,\ S_{i,j}^k=\mathrm{MR\text{-}M} \\ 
        &4,\ S_{i,j}^k=\mathrm{DR\text{-}M}\\ 
    \end{aligned}
  \right. .
\end{equation}

And for each data qubit pair $(d_{i,j},d_{p,q}) \in \mathcal{G}$, we evaluate two measurement strategies: (1) Joint parity measurement with unified cost $C_\text{joint}(S_{i,j}^k)$
\begin{equation}\label{eq-local cost2}
  C_\text{joint}(S_{i,j}^k)=
  \left\{
    \begin{aligned}
        &\min\{\epsilon_m(d_{i,j}),\epsilon_m(d_{p,q})\}+\mathcal{N}(S_{i,j}^k)\cdot\epsilon_g \\
          &\quad \quad \quad \quad \quad \quad \quad \quad \quad \quad , \ S_{i,j}^k=\mathrm{DR\text{-}PM} \\
        &\epsilon_m(m_{k,l}^\sigma)+\mathcal{N}(S_{i,j}^k)\cdot\epsilon_g,\ S_{i,j}^k=\mathrm{MR\text{-}PM} \\ 
    \end{aligned}
  \right. ; 
\end{equation} 
(2) Independent single-qubit measurements with summed costs $(C_{\text{ind}}(S_{i,j}^k)+C_{\text{ind}}(S_{p,q}^{k'}))$. Here the number $\mathcal{N}(S_{i,j}^k)$ for $\mathrm{DR\text{-}PM}$ and $\mathrm{MR\text{-}PM}$ is
\begin{equation}\label{eq-gate_num2}
  \mathcal{N}(S_{i,j}^k)=
  \left\{
    \begin{aligned}
        &2,\ S_{i,j}^k=\mathrm{MR\text{-}PM} \\
        &3,\ S_{i,j}^k=\mathrm{DR\text{-}PM} \\  
    \end{aligned}
  \right. .
\end{equation}
The optimal modality is selected by:  
\begin{equation}
\text{argmin}\left(C_{\text{joint}}(S_{i,j}^k),\ (C_{\text{ind}}(S_{i,j}^k)+C_{\text{ind}}(S_{p,q}^{k'}))\right),  
\end{equation}  
enabling adaptive choice between joint or independent modalities for each pair.

In constructing measurement circuits, we first verify the availability of all qubits involved in the selected modality. If any qubit is occupied, the measurement is scheduled sequentially within the available tick. For simplicity, we omit the differences between two-qubit gates applied to different qubits. However, when implementing on noisy quantum devices, one can adjust the latter part $\mathcal{N}(\cdot)\cdot \epsilon_g$ of $\mathcal{C}(S_{i,j}^k)$ in Eq.~\ref{eq-local cost1} and Eq.~\ref{eq-local cost2} to account for CNOT gate errors at different locations. 

\subsection{Numerical simulations}
We investigate the MS-local protocol using numerical simulations of the rotated surface code with the Python library Stim~\cite{stim_library}. In Fig.~\ref{fig4}, we conduct quantum memory experiments where a logical qubit initialized in $|0\rangle_L$ undergoes $r$ stabilizer measurement rounds and final readout. This setup evaluates the code's long-term error correction capability under persistent noise.
Simulations integrate error metrics from four noisy quantum devices: {\textit{ibm-ithaca}}~\cite{Morvan2024Phase}, {\textit{Sycamore}}~\cite{Arute2019Quantum}, and {\textit{zuchongzhi-2.0}}~\cite{Wu2021Strong} and {\textit{zuchongzhi-3.0}}~\cite{Gao2025Establishing}. These devices exhibit distinct error profiles. {\textit{zuchongzhi-2.0}} and {\textit{ibm-ithaca}} show a pronounced imbalance between gate and measurement error rates ($\epsilon_g << \epsilon_m$), while {\textit{Sycamore}} and {\textit{zuchongzhi-3.0}} have narrower gaps. 
In addition to gate and measurement noise, we incorporated depolarizing noise affecting data qubits both during syndrome measurement rounds and while idle during scheduling ticks. We primarily present results for {\textit{ibm-ithaca}} and {\textit{Sycamore}}. Details of the error profiles, experimental settings, and results for other devices are provided in Supplemental materials.

We first explore the underlying mechanism of MS by isolating the impacts of data-qubit measurement errors through controlled MER scaling. Here, the data-qubit MER scaling parameter $\alpha$ represents synthetic error suppression: $\alpha=1.0$ corresponds to the original hardware MER, while $\alpha=0.5$ simulates a 50\% MER reduction. In Fig.~\ref{fig4} \textbf{a}, fixing the measurement-qubit MER while selectively increasing the data-qubit MER in a $d=5$ surface code reveals linear logical error rate growth (slope = $0.02 \pm 0.0013$), directly linking data-qubit MER to the logical error rate.
The efficacy of MS-local scheduling is quantified by comparing it against data-qubit MER-scaled baselines in Fig.~\ref{fig4} \textbf{b}-\textbf{c}. Remarkably, MS-local achieves logical error rate equivalent to at least a 50\% reduction ($\alpha \leq 0.5$) across all code distances ($d=$3 to 11) for both two devices. This effectively reduces the effective data-qubit MER, enhancing logical readout performance to a level comparable to halving the native MER of data qubits. 

In Fig.~\ref{fig4} \textbf{d}, the logical error rate ratio (Original/MS-local) shows that the performance gain diminishes as the ratio $\epsilon_g / \epsilon_m$ increases, reflecting the trade-off between MER and GER in measurement transfer.
Specifically, on \textit{ibm-ithaca} ($\epsilon_g/\epsilon_m=0.2125$), MS-local achieves a 34\% error reduction, while on \textit{Sycamore} ($\epsilon_g/\epsilon_m=0.3083$), the reduction is 7.5\%.  
Analysis of effective code distance (ECD) is given in Fig.~\ref{fig4} \textbf{e}-\textbf{f}. The ECD of Original equals code distance $d$. For other cases, ECD is calculated using the formula error$=a(p/p_\mathrm{th})^{\text{ECD}/2}$, with parameters $a$ and $p/p_\mathrm{th}$ fitted from Original's error-$d$ data. Results indicate that MS-local achieves an average effective code distance of $d_\mathrm{eff}=d+3.32$ for \textit{ibm-ithaca} and $d_\mathrm{eff}=d+0.49$ for \text{Sycamore}. This corresponds to a physical qubit reduction of approximately 51\% for \textit{ibm-ithaca} and 9\% for \textit{Sycamore} at $d=11$. 
In addition, as demonstrated in Fig.~\ref{fig4} \textbf{h-i}, the logical error rate has a non-monotonic dependence on measurement error heterogeneity. The error initially decreases as the standard deviation approaches near-unity scale, but rebounds under excessive noise dispersion. MS-local demonstrates enhanced error suppression during the descending phase and mitigated degradation in the ascending phase, owing to its modality-aware measurement transfer that exploits qubit-specific readout accuracy. Consequently, the performance gap widens significantly with rising noise variability. 

These experimental results collectively demonstrate that MS-local significantly enhances logical readout performance by effectively modulating data-qubit MER, reducing logical error rate and improving effective code distance across platforms.

\section{Reinforcement learning-guided measurement scheduling with temporal constraints}\label{sec4} 
\begin{figure*}[t]
  \centering
  {\includegraphics[width=0.75\textwidth]{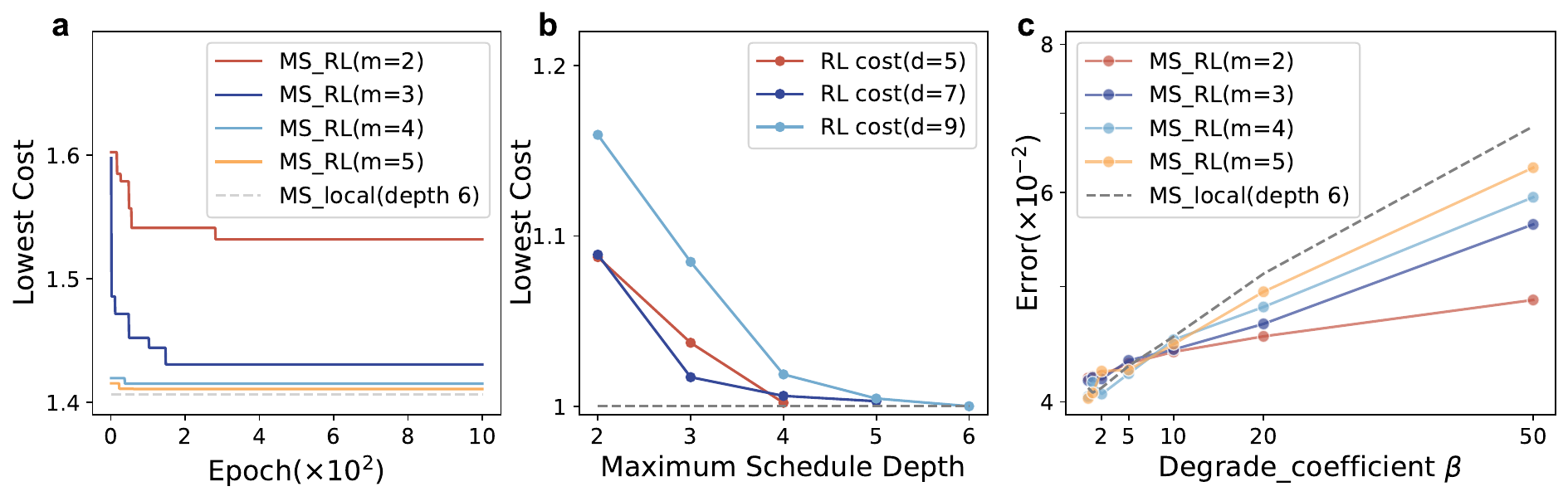}} 
  \caption{{\textbf{The performance of Reinforcement learning-guided Measurement Scheduling (MS-RL) with temporal constraints (maximum schedule depth $m$) across quantum devices.}  
  Simulation experiments evaluate logical error rate reduction of MS-RL in comparison with MS with local greedy assignment (MS-local) with device-calibrated measurement error data obtained from \textit{ibm-ithaca}.
  \textbf{a} Lowest cost of MS-RL with depth constraints $m=2$ to $5$ over $10^3$ training epochs with MS-local (gray dashed) as baseline at scheduling depth 6 for code distance $d=7$. \textbf{b} The lowest cost varying with different maximum schedule depth $m$ with code distance $d=5,7,9$. \textbf{c} Logical error rate for $d=7$ under MS-RL ($m=2$ to $5$) and MS-local with increasing degrade coefficient $\beta$, representing the data qubit degradation rate during idle scheduling waits (coefficient 1 denoting the true degradation rate).}}
  \label{fig5}
\end{figure*}

We now extend our MS framework to incorporate temporal constraints, thereby advancing towards a more realistic model by accounting for the delays induced by modality conflicts, which amplify the impact of decoherence. This extension reveals two critical challenges. First, extended readout durations due to these conflicts increase the temporal depth of the syndrome history, complicating the union-find graph used in decoding. Second, asynchronous measurements cause qubit-specific idling, leading to coherence-time-dependent error accumulation. To address these challenges, we develop a spatio-temporal co-optimization strategy that integrates temporal dynamics into MS.

To formalize the temporal constraints within our MS framework, we define the schedule depth. Let $m \in \mathbb{Z}^{+}$ denote the maximum allowable schedule depth, which is defined as the number of discrete temporal ticks required to resolve all measurement conflicts under the scheduling policy. With this formalization, we now introduce two new decision variables: 
\begin{itemize}
  \item $\{x_{i,j}^{k,t}\} \in \{0, 1\}$ indicates whether modality $S_{i,j}^k$ is scheduled at time tick $t$.
  \item $\tau \in \mathbb{Z}^{+}$ denotes the current total temporal ticks (queue length).
\end{itemize} 
Then the integer programming objective in Eq.~\ref{eq-inpro-formulation} can be reformulated to
\begin{equation}\label{eq-tempo-formulation}
  \underset{{x_{i,j}^{k,t}}}{\text{minimize}} \sum_{i,j} \sum_{k=1}^{K_{ij}}\sum_{t=1}^\tau x_{i,j}^{k,t} \cdot \mathcal{C}(S_{i,j}^k),
\end{equation}
\textit{subject to:}
\begin{itemize}
  \item Modality Selection:
  \begin{equation}
    \sum_{k=1}^{K_{ij}} \sum_{t=1}^m x_{i,j}^{k,t} = 1, \forall d_{i,j} \in \mathcal{D}.
  \end{equation}
   
  \item Conflict-Free Scheduling: 
  \begin{equation}
    \sum_{\substack{(p,q,k') \\ \mathcal{R}_{i,j}^k \cap \mathcal{R}_{p,q}^{k'} \neq \varnothing}} x_{p,q}^{k',t} \leq 1 - x_{i,j}^{k,t}, \forall i,j,k,t.
  \end{equation}
  \item Queue Length Bound: 
  \begin{equation}
    \tau \leq m, \quad \text{where} \  \tau = \max\left\{ t \mid \exists x_{i,j}^{k,t} = 1 \right\}.
  \end{equation} 
\end{itemize}
Directly applying classical optimization methods, such as integer programming and conflict-directed backtracking, to handle time-dependent scheduling constraints is computationally intractable due to their prohibitive complexity $\mathcal{O}(e^{|\mathcal{D}|})$. This issue is exacerbated in fault-tolerant architectures, where the number of defect pairs $|\mathcal{D}|$ scales quadratically with the number of physical qubits $N$ ($|\mathcal{D}| \sim N^2$).

\subsection{Reinforcement learning-guided measurement scheduling}
To address this scaling challenge, we develop an RL-guided measurement scheduling framework (MS-RL) that jointly optimizes measurement redistribution under time constraints, thereby circumventing combinatorial explosion. 
The framework employs pointer-based strategy optimization with adaptive reward mechanisms. Each data qubit maintains a sorted sequence of measurement modalities $S_i = \{s_1^i, s_2^i, ..., s_k^i\}$ filtered with $C(s_j^i) \leq C_{\mathrm{D}\text{-}\mathrm{M}}$, with strategies ordered by ascending costs. After each action selected by the agent, the environment schedules the chosen modalities for all data qubits and returns the total queue length $\tau$.
The dual optimization process minimizes measurement costs through discrete pointer adjustments while regulating system stability via time-dependent reward shaping.

\subsubsection{Action Space Configuration}
The agent iteratively adjusts these sequences through discrete pointer moves, guided by adaptive reward mechanisms. Each action $(i,\Delta p_i)$ selects a target qubit $i$ and shifts its sequence pointer $p_i$ by the pointer adjustment $\Delta p_i=\pm 1$ unit. The environment dynamically enforces boundary constraints by disabling moves beyond precomputed modality sequence lengths $\{k_i\}_i$. Consequently, the available action set at time $t$ is determined as:
\begin{equation}
  \mathcal{A}_t^{\text{valid}} = \bigcup_{i=1}^N \{\Delta p_i | (p_i + \Delta p_i) \in [1,k_i]\}
  \end{equation}
This constraint reduces the action space from exponential complexity to linear $2\times N_d$ options, where is $N_d$ the data-qubit count. By converting global strategy selection into local pointer navigation (qubit selection and direction choice), the search process becomes equivalent to traversing constrained grid paths.

\subsubsection{Reward Function Design}
The reward function in our MS-RL framework is designed to integrate three dynamically weighted components that guide the reinforcement learning process. Specifically, $\Delta Q^-$ is the amount by which the queue length decreases, and $\Delta Q^+$ is the amount by which the queue length increases. The reward function is defined as:
\begin{equation}
  r_t = \underbrace{\alpha(t)\Delta Q^-/m}_{\text{queue reward}} - \underbrace{\beta(t)\Delta Q^+/m}_{\text{penalty term}} - \underbrace{\gamma\sum \mathcal{C}(s_{p_i}^i)}_{\text{cost term}},
\end{equation}
where $\alpha(t)$ and $\beta(t)$ are time-varying weights that implement a scheduling mechanism to balance the immediate and long-term effects of actions on the queue length. The term $\gamma\sum \mathcal{C}(s_{p_i}^i)$ represents the cost associated with the chosen measurement modalities. 

The advantage of this formula lies in its flexibility to handle the temporary increase in queue length caused by individual actions with a delay. For simplification, the dynamic weights $\alpha(t)$ and $\beta(t)$ can be replaced with fixed rewards, provided that the convergence of the learning process is ensured.

\subsection{Numerical Simulations}
We evaluate MS-RL's temporal constraint performance using \textit{ibm-ithaca}-calibrated measurement error data. 
As shown in Fig.~\ref{fig5} \textbf{a}, MS-RL achieves rapid convergence within 300 training epochs for all tested maximum schedule depths ($m=2$ to $5$). Configurations with $m \geq 3$ stabilize within 200 epochs, indicating practical scalability for real-device deployment. Meanwhile, Fig.~\ref{fig5} \textbf{b} reveals significant optimization potential in the cost-depth tradeoff: increasing the maximum schedule depth $m$ drives substantial cost reduction, though with diminishing returns beyond $m=4$ as the system approaches near-optimal scheduling efficiency.

Furthermore, Fig.~\ref{fig5} \textbf{c} demonstrates MS-RL's temporal scheduling robustness through qubit decay simulations, where we scale data qubit degradation rate $\beta$ during idle waits ($\beta=1$ reflects real-world conditions). MS-RL exhibits superior robustness against qubit decay compared to MS-local as $\beta$ increase. Notably, scheduler depth 
$m$ exerts stronger control over error suppression at elevated $\beta$. While shallow $m$ schedulers initially lag behind deeper counterparts, they outperform deeper schedulers beyond $\beta\approx 8$ for code distance $d=7$, exhibiting a discrete threshold response. This $\beta$-dependent dominance defines optimal $m$ selection regimes in MS-RL, revealing critical competition between MS-induced idling damage and temporal constraint efficiency. Crucially, such threshold dynamics prove critical for devices with coherence limitations where extended idling during logical measurement cycles catastrophically amplifies error accumulation. 

\section{Conclusion}
In this work, we present a measurement scheduling (MS) protocol that suppresses logical readout errors through measurement transfer modalities, which serve as the scheduling units. These modalities are designed to utilize shallow entangled circuits to transfer measurement tasks, introducing a trade-off between introduced gate errors and measurement errors. We first introduce two basic measurement transfer modalities and then design surface code-specific modalities tailored to the logical readout tasks and topological structure of surface code. These modalities enable the protocol to dynamically route measurements from error-prone qubits to stable nodes through adaptive resource allocation, resolving conflicts via real-time error metrics and topological criticality. By integrating reinforcement learning, the framework further optimizes temporal constraints imposed by decoherence limits and error correction deadlines without compromising computational efficiency. 

Numerical simulations validate the operational mechanism of the MS protocol, demonstrating its equivalence to reducing the effective measurement error rate (MER) of data qubits. Both local MS and reinforcement learning (RL)-guided MS outperform the original direct measurement strategy across code distances $d=3$ to $11$, achieving an average reduction in logical error rate of up to 34\%.  Further analysis reveals two critical factors influencing MS performance: the gate-to-measurement error ratio and the heterogeneity of measurement errors. The protocol is most effective in noisy devices where measurement noise dominates gate noise, while maintaining robustness against heterogeneous error distributions. Additionally, we establish the scalability of RL-guided MS, successfully simulating its application up to $d=9$, thereby highlighting its potential as a versatile and extensible solution for large-scale quantum systems. 

The MS protocol effectively reduces logical error rates, enhancing the reliability of quantum error correction and advancing large-scale quantum computing.
However, it may cause extra time consumption due to scheduling ticks exceeding one when conflicts occur. Yet, numerical simulations suggest that the MS protocol can achieve the same logical qubit fidelity with fewer encoding qubits, reducing encoding and stabilizer measurement time, which may partially offset its additional time consumption.
Future work can explore extending the MS protocol to other readout scenarios, as our modality design is inherently flexible and compatible with any qubit topology. Additionally, incorporating spatially resolved two-qubit gate errors into the scheduling process could further enhance practicality. This approach would effectively weight the edges in the integer programming optimization, better aligning the protocol with real-world error landscapes.

\bibliography{b}

\begin{thebibliography}{33}%
\makeatletter
\providecommand \@ifxundefined [1]{%
 \@ifx{#1\undefined}
}%
\providecommand \@ifnum [1]{%
 \ifnum #1\expandafter \@firstoftwo
 \else \expandafter \@secondoftwo
 \fi
}%
\providecommand \@ifx [1]{%
 \ifx #1\expandafter \@firstoftwo
 \else \expandafter \@secondoftwo
 \fi
}%
\providecommand \natexlab [1]{#1}%
\providecommand \enquote  [1]{``#1''}%
\providecommand \bibnamefont  [1]{#1}%
\providecommand \bibfnamefont [1]{#1}%
\providecommand \citenamefont [1]{#1}%
\providecommand \href@noop [0]{\@secondoftwo}%
\providecommand \href [0]{\begingroup \@sanitize@url \@href}%
\providecommand \@href[1]{\@@startlink{#1}\@@href}%
\providecommand \@@href[1]{\endgroup#1\@@endlink}%
\providecommand \@sanitize@url [0]{\catcode `\\12\catcode `\$12\catcode
  `\&12\catcode `\#12\catcode `\^12\catcode `\_12\catcode `\%12\relax}%
\providecommand \@@startlink[1]{}%
\providecommand \@@endlink[0]{}%
\providecommand \url  [0]{\begingroup\@sanitize@url \@url }%
\providecommand \@url [1]{\endgroup\@href {#1}{\urlprefix }}%
\providecommand \urlprefix  [0]{URL }%
\providecommand \Eprint [0]{\href }%
\providecommand \doibase [0]{http://dx.doi.org/}%
\providecommand \selectlanguage [0]{\@gobble}%
\providecommand \bibinfo  [0]{\@secondoftwo}%
\providecommand \bibfield  [0]{\@secondoftwo}%
\providecommand \translation [1]{[#1]}%
\providecommand \BibitemOpen [0]{}%
\providecommand \bibitemStop [0]{}%
\providecommand \bibitemNoStop [0]{.\EOS\space}%
\providecommand \EOS [0]{\spacefactor3000\relax}%
\providecommand \BibitemShut  [1]{\csname bibitem#1\endcsname}%
\let\auto@bib@innerbib\@empty
\bibitem [{\citenamefont {Cory}\ \emph {et~al.}(1998)\citenamefont {Cory},
  \citenamefont {Price}, \citenamefont {Maas}, \citenamefont {Knill},
  \citenamefont {Laflamme}, \citenamefont {Zurek}, \citenamefont {Havel},\ and\
  \citenamefont {Somaroo}}]{Cory1998Experimental}%
  \BibitemOpen
  \bibfield  {author} {\bibinfo {author} {\bibfnamefont {D.~G.}\ \bibnamefont
  {Cory}}, \bibinfo {author} {\bibfnamefont {M.~D.}\ \bibnamefont {Price}},
  \bibinfo {author} {\bibfnamefont {W.}~\bibnamefont {Maas}}, \bibinfo {author}
  {\bibfnamefont {E.}~\bibnamefont {Knill}}, \bibinfo {author} {\bibfnamefont
  {R.}~\bibnamefont {Laflamme}}, \bibinfo {author} {\bibfnamefont {W.~H.}\
  \bibnamefont {Zurek}}, \bibinfo {author} {\bibfnamefont {T.~F.}\ \bibnamefont
  {Havel}}, \ and\ \bibinfo {author} {\bibfnamefont {S.~S.}\ \bibnamefont
  {Somaroo}},\ }\href {\doibase 10.1103/PhysRevLett.81.2152} {\bibfield
  {journal} {\bibinfo  {journal} {Phys. Rev. Lett.}\ }\textbf {\bibinfo
  {volume} {81}},\ \bibinfo {pages} {2152} (\bibinfo {year}
  {1998})}\BibitemShut {NoStop}%
\bibitem [{\citenamefont {Knill}\ \emph {et~al.}(2000)\citenamefont {Knill},
  \citenamefont {Laflamme},\ and\ \citenamefont {Viola}}]{Knill2000Theory}%
  \BibitemOpen
  \bibfield  {author} {\bibinfo {author} {\bibfnamefont {E.}~\bibnamefont
  {Knill}}, \bibinfo {author} {\bibfnamefont {R.}~\bibnamefont {Laflamme}}, \
  and\ \bibinfo {author} {\bibfnamefont {L.}~\bibnamefont {Viola}},\ }\href
  {\doibase 10.1103/PhysRevLett.84.2525} {\bibfield  {journal} {\bibinfo
  {journal} {Phys. Rev. Lett.}\ }\textbf {\bibinfo {volume} {84}},\ \bibinfo
  {pages} {2525} (\bibinfo {year} {2000})}\BibitemShut {NoStop}%
\bibitem [{\citenamefont {Chiaverini}\ \emph {et~al.}(2004)\citenamefont
  {Chiaverini}, \citenamefont {Leibfried}, \citenamefont {Schaetz},
  \citenamefont {Barrett}, \citenamefont {Blakestad}, \citenamefont {Britton},
  \citenamefont {Itano}, \citenamefont {Jost}, \citenamefont {Knill},
  \citenamefont {Langer}, \citenamefont {Ozeri},\ and\ \citenamefont
  {Wineland}}]{Chiaverini2004Realization}%
  \BibitemOpen
  \bibfield  {author} {\bibinfo {author} {\bibfnamefont {J.}~\bibnamefont
  {Chiaverini}}, \bibinfo {author} {\bibfnamefont {D.}~\bibnamefont
  {Leibfried}}, \bibinfo {author} {\bibfnamefont {T.}~\bibnamefont {Schaetz}},
  \bibinfo {author} {\bibfnamefont {M.~D.}\ \bibnamefont {Barrett}}, \bibinfo
  {author} {\bibfnamefont {R.~B.}\ \bibnamefont {Blakestad}}, \bibinfo {author}
  {\bibfnamefont {J.}~\bibnamefont {Britton}}, \bibinfo {author} {\bibfnamefont
  {W.~M.}\ \bibnamefont {Itano}}, \bibinfo {author} {\bibfnamefont {J.~D.}\
  \bibnamefont {Jost}}, \bibinfo {author} {\bibfnamefont {E.}~\bibnamefont
  {Knill}}, \bibinfo {author} {\bibfnamefont {C.}~\bibnamefont {Langer}},
  \bibinfo {author} {\bibfnamefont {R.}~\bibnamefont {Ozeri}}, \ and\ \bibinfo
  {author} {\bibfnamefont {D.~J.}\ \bibnamefont {Wineland}},\ }\href {\doibase
  10.1038/nature03074} {\bibfield  {journal} {\bibinfo  {journal} {Nature}\
  }\textbf {\bibinfo {volume} {432}},\ \bibinfo {pages} {602} (\bibinfo {year}
  {2004})}\BibitemShut {NoStop}%
\bibitem [{\citenamefont {Reichardt}\ \emph {et~al.}(2024)\citenamefont
  {Reichardt} \emph {et~al.}}]{Reichardt2024Logical}%
  \BibitemOpen
  \bibfield  {author} {\bibinfo {author} {\bibfnamefont {B.~W.}\ \bibnamefont
  {Reichardt}} \emph {et~al.},\ }\href {https://arxiv.org/abs/2411.11822}
  {\bibfield  {journal} {\bibinfo  {journal} {arXiv:2411.11822}\ } (\bibinfo
  {year} {2024})}\BibitemShut {NoStop}%
\bibitem [{\citenamefont {Paetznick}\ \emph {et~al.}(2024)\citenamefont
  {Paetznick} \emph {et~al.}}]{Paetznick2024Demonstration}%
  \BibitemOpen
  \bibfield  {author} {\bibinfo {author} {\bibfnamefont {A.}~\bibnamefont
  {Paetznick}} \emph {et~al.},\ }\href {https://arxiv.org/abs/2404.02280}
  {\bibfield  {journal} {\bibinfo  {journal} {arXiv:2404.02280}\ } (\bibinfo
  {year} {2024})}\BibitemShut {NoStop}%
\bibitem [{\citenamefont {Acharya}\ \emph {et~al.}(2024)\citenamefont {Acharya}
  \emph {et~al.}}]{Google2024Quantum}%
  \BibitemOpen
  \bibfield  {author} {\bibinfo {author} {\bibfnamefont {R.}~\bibnamefont
  {Acharya}} \emph {et~al.},\ }\href
  {https://api.semanticscholar.org/CorpusID:274608118} {\bibfield  {journal}
  {\bibinfo  {journal} {Nature}\ } (\bibinfo {year} {2024})}\BibitemShut
  {NoStop}%
\bibitem [{\citenamefont {Acharya}\ \emph {et~al.}(2023)\citenamefont {Acharya}
  \emph {et~al.}}]{Acharya2023Suppressing}%
  \BibitemOpen
  \bibfield  {author} {\bibinfo {author} {\bibfnamefont {R.}~\bibnamefont
  {Acharya}} \emph {et~al.},\ }\href {\doibase 10.1038/s41586-022-05434-1}
  {\bibfield  {journal} {\bibinfo  {journal} {Nature}\ }\textbf {\bibinfo
  {volume} {614}},\ \bibinfo {pages} {676} (\bibinfo {year}
  {2023})}\BibitemShut {NoStop}%
\bibitem [{\citenamefont {Fowler}\ \emph {et~al.}(2012)\citenamefont {Fowler},
  \citenamefont {Mariantoni}, \citenamefont {Martinis},\ and\ \citenamefont
  {Cleland}}]{Fowler2012Surface}%
  \BibitemOpen
  \bibfield  {author} {\bibinfo {author} {\bibfnamefont {A.~G.}\ \bibnamefont
  {Fowler}}, \bibinfo {author} {\bibfnamefont {M.}~\bibnamefont {Mariantoni}},
  \bibinfo {author} {\bibfnamefont {J.~M.}\ \bibnamefont {Martinis}}, \ and\
  \bibinfo {author} {\bibfnamefont {A.~N.}\ \bibnamefont {Cleland}},\ }\href
  {\doibase 10.1103/PhysRevA.86.032324} {\bibfield  {journal} {\bibinfo
  {journal} {Phys. Rev. A}\ }\textbf {\bibinfo {volume} {86}},\ \bibinfo
  {pages} {032324} (\bibinfo {year} {2012})}\BibitemShut {NoStop}%
\bibitem [{\citenamefont {Zhao}\ \emph {et~al.}(2022)\citenamefont {Zhao} \emph
  {et~al.}}]{Zhao2022Realization}%
  \BibitemOpen
  \bibfield  {author} {\bibinfo {author} {\bibfnamefont {Y.}~\bibnamefont
  {Zhao}} \emph {et~al.},\ }\href {\doibase 10.1103/PhysRevLett.129.030501}
  {\bibfield  {journal} {\bibinfo  {journal} {Phys. Rev. Lett.}\ }\textbf
  {\bibinfo {volume} {129}},\ \bibinfo {pages} {030501} (\bibinfo {year}
  {2022})}\BibitemShut {NoStop}%
\bibitem [{\citenamefont {Fowler}\ \emph {et~al.}(2010)\citenamefont {Fowler}
  \emph {et~al.}}]{Fowler2010Surface}%
  \BibitemOpen
  \bibfield  {author} {\bibinfo {author} {\bibfnamefont {A.~G.}\ \bibnamefont
  {Fowler}} \emph {et~al.},\ }\href {\doibase 10.1103/PhysRevLett.104.180503}
  {\bibfield  {journal} {\bibinfo  {journal} {Phys. Rev. Lett.}\ }\textbf
  {\bibinfo {volume} {104}},\ \bibinfo {pages} {180503} (\bibinfo {year}
  {2010})}\BibitemShut {NoStop}%
\bibitem [{\citenamefont {Horsman}\ \emph {et~al.}(2012)\citenamefont {Horsman}
  \emph {et~al.}}]{Horsman2012surgery}%
  \BibitemOpen
  \bibfield  {author} {\bibinfo {author} {\bibfnamefont {D.}~\bibnamefont
  {Horsman}} \emph {et~al.},\ }\href {\doibase 10.1088/1367-2630/14/12/123011}
  {\bibfield  {journal} {\bibinfo  {journal} {New Journal of Physics}\ }\textbf
  {\bibinfo {volume} {14}},\ \bibinfo {pages} {123011} (\bibinfo {year}
  {2012})}\BibitemShut {NoStop}%
\bibitem [{\citenamefont {Foxen}\ \emph {et~al.}(2020)\citenamefont {Foxen}
  \emph {et~al.}}]{Foxen20202020Demonstrating}%
  \BibitemOpen
  \bibfield  {author} {\bibinfo {author} {\bibfnamefont {B.}~\bibnamefont
  {Foxen}} \emph {et~al.} (\bibinfo {collaboration} {Google AI Quantum}),\
  }\href {\doibase 10.1103/PhysRevLett.125.120504} {\bibfield  {journal}
  {\bibinfo  {journal} {Phys. Rev. Lett.}\ }\textbf {\bibinfo {volume} {125}},\
  \bibinfo {pages} {120504} (\bibinfo {year} {2020})}\BibitemShut {NoStop}%
\bibitem [{\citenamefont {Xu}\ \emph {et~al.}(2020)\citenamefont {Xu} \emph
  {et~al.}}]{Yuan2020High}%
  \BibitemOpen
  \bibfield  {author} {\bibinfo {author} {\bibfnamefont {Y.}~\bibnamefont {Xu}}
  \emph {et~al.},\ }\href {\doibase 10.1103/PhysRevLett.125.240503} {\bibfield
  {journal} {\bibinfo  {journal} {Phys. Rev. Lett.}\ }\textbf {\bibinfo
  {volume} {125}},\ \bibinfo {pages} {240503} (\bibinfo {year}
  {2020})}\BibitemShut {NoStop}%
\bibitem [{\citenamefont {Neg\^{\i}rneac}\ \emph {et~al.}(2021)\citenamefont
  {Neg\^{\i}rneac} \emph {et~al.}}]{V2021High}%
  \BibitemOpen
  \bibfield  {author} {\bibinfo {author} {\bibfnamefont {V.}~\bibnamefont
  {Neg\^{\i}rneac}} \emph {et~al.},\ }\href {\doibase
  10.1103/PhysRevLett.126.220502} {\bibfield  {journal} {\bibinfo  {journal}
  {Phys. Rev. Lett.}\ }\textbf {\bibinfo {volume} {126}},\ \bibinfo {pages}
  {220502} (\bibinfo {year} {2021})}\BibitemShut {NoStop}%
\bibitem [{\citenamefont {Sung}\ \emph {et~al.}(2021)\citenamefont {Sung} \emph
  {et~al.}}]{Sung2021Realization}%
  \BibitemOpen
  \bibfield  {author} {\bibinfo {author} {\bibfnamefont {Y.}~\bibnamefont
  {Sung}} \emph {et~al.},\ }\href {\doibase 10.1103/PhysRevX.11.021058}
  {\bibfield  {journal} {\bibinfo  {journal} {Phys. Rev. X}\ }\textbf {\bibinfo
  {volume} {11}},\ \bibinfo {pages} {021058} (\bibinfo {year}
  {2021})}\BibitemShut {NoStop}%
\bibitem [{\citenamefont {Walter}\ \emph {et~al.}(2017)\citenamefont {Walter}
  \emph {et~al.}}]{Walter2017Rapid}%
  \BibitemOpen
  \bibfield  {author} {\bibinfo {author} {\bibfnamefont {T.}~\bibnamefont
  {Walter}} \emph {et~al.},\ }\href {\doibase 10.1103/PhysRevApplied.7.054020}
  {\bibfield  {journal} {\bibinfo  {journal} {Phys. Rev. Appl.}\ }\textbf
  {\bibinfo {volume} {7}},\ \bibinfo {pages} {054020} (\bibinfo {year}
  {2017})}\BibitemShut {NoStop}%
\bibitem [{\citenamefont {Bultink}\ \emph {et~al.}(2020)\citenamefont {Bultink}
  \emph {et~al.}}]{Bultink2020Protecting}%
  \BibitemOpen
  \bibfield  {author} {\bibinfo {author} {\bibfnamefont {C.~C.}\ \bibnamefont
  {Bultink}} \emph {et~al.},\ }\href {\doibase 10.1126/sciadv.aay3050}
  {\bibfield  {journal} {\bibinfo  {journal} {Science Advances}\ }\textbf
  {\bibinfo {volume} {6}},\ \bibinfo {pages} {eaay3050} (\bibinfo {year}
  {2020})}\BibitemShut {NoStop}%
\bibitem [{\citenamefont {Linke}\ \emph {et~al.}(2017)\citenamefont {Linke}
  \emph {et~al.}}]{Linke2017FaultFault}%
  \BibitemOpen
  \bibfield  {author} {\bibinfo {author} {\bibfnamefont {N.~M.}\ \bibnamefont
  {Linke}} \emph {et~al.},\ }\href {\doibase 10.1126/sciadv.1701074} {\bibfield
   {journal} {\bibinfo  {journal} {Science Advances}\ }\textbf {\bibinfo
  {volume} {3}},\ \bibinfo {pages} {e1701074} (\bibinfo {year}
  {2017})}\BibitemShut {NoStop}%
\bibitem [{\citenamefont {Chen}\ \emph {et~al.}(2023)\citenamefont {Chen} \emph
  {et~al.}}]{Chen2023Transmon}%
  \BibitemOpen
  \bibfield  {author} {\bibinfo {author} {\bibfnamefont {L.}~\bibnamefont
  {Chen}} \emph {et~al.},\ }\href {\doibase 10.1038/s41534-023-00689-6}
  {\bibfield  {journal} {\bibinfo  {journal} {npj Quantum Information}\
  }\textbf {\bibinfo {volume} {9}},\ \bibinfo {pages} {26} (\bibinfo {year}
  {2023})}\BibitemShut {NoStop}%
\bibitem [{\citenamefont {Pino}\ \emph {et~al.}(2021)\citenamefont {Pino} \emph
  {et~al.}}]{Pino2021Demonstration}%
  \BibitemOpen
  \bibfield  {author} {\bibinfo {author} {\bibfnamefont {J.~M.}\ \bibnamefont
  {Pino}} \emph {et~al.},\ }\href {\doibase 10.1038/s41586-021-03318-4}
  {\bibfield  {journal} {\bibinfo  {journal} {Nature}\ }\textbf {\bibinfo
  {volume} {592}},\ \bibinfo {pages} {209} (\bibinfo {year}
  {2021})}\BibitemShut {NoStop}%
\bibitem [{\citenamefont {Tantivasadakarn}\ \emph {et~al.}(2023)\citenamefont
  {Tantivasadakarn}, \citenamefont {Vishwanath},\ and\ \citenamefont
  {Verresen}}]{Tantivasadakarn2023Hierarchy}%
  \BibitemOpen
  \bibfield  {author} {\bibinfo {author} {\bibfnamefont {N.}~\bibnamefont
  {Tantivasadakarn}}, \bibinfo {author} {\bibfnamefont {A.}~\bibnamefont
  {Vishwanath}}, \ and\ \bibinfo {author} {\bibfnamefont {R.}~\bibnamefont
  {Verresen}},\ }\href {\doibase 10.1103/PRXQuantum.4.020339} {\bibfield
  {journal} {\bibinfo  {journal} {PRX Quantum}\ }\textbf {\bibinfo {volume}
  {4}},\ \bibinfo {pages} {020339} (\bibinfo {year} {2023})}\BibitemShut
  {NoStop}%
\bibitem [{\citenamefont {Smith}\ \emph {et~al.}(2023)\citenamefont {Smith}
  \emph {et~al.}}]{Smith2023Scaling}%
  \BibitemOpen
  \bibfield  {author} {\bibinfo {author} {\bibfnamefont {K.~N.}\ \bibnamefont
  {Smith}} \emph {et~al.},\ }in\ \href {\doibase 10.1109/MICRO56248.2022.00078}
  {\emph {\bibinfo {booktitle} {Proceedings of the 55th Annual IEEE/ACM
  International Symposium on Microarchitecture}}},\ \bibinfo {series and
  number} {MICRO '22}\ (\bibinfo  {publisher} {IEEE Press},\ \bibinfo {year}
  {2023})\ p.\ \bibinfo {pages} {1092–1109}\BibitemShut {NoStop}%
\bibitem [{\citenamefont {Mnih}\ \emph {et~al.}(2015)\citenamefont {Mnih} \emph
  {et~al.}}]{Mnih2015Human}%
  \BibitemOpen
  \bibfield  {author} {\bibinfo {author} {\bibfnamefont {V.}~\bibnamefont
  {Mnih}} \emph {et~al.},\ }\href {\doibase 10.1038/nature14236} {\bibfield
  {journal} {\bibinfo  {journal} {Nature}\ }\textbf {\bibinfo {volume} {518}},\
  \bibinfo {pages} {529} (\bibinfo {year} {2015})}\BibitemShut {NoStop}%
\bibitem [{\citenamefont {Silver}\ \emph {et~al.}(2016)\citenamefont {Silver}
  \emph {et~al.}}]{Silver2016Mastering}%
  \BibitemOpen
  \bibfield  {author} {\bibinfo {author} {\bibfnamefont {D.}~\bibnamefont
  {Silver}} \emph {et~al.},\ }\href {\doibase 10.1038/nature16961} {\bibfield
  {journal} {\bibinfo  {journal} {Nature}\ }\textbf {\bibinfo {volume} {529}},\
  \bibinfo {pages} {484} (\bibinfo {year} {2016})}\BibitemShut {NoStop}%
\bibitem [{\citenamefont {Silver}\ \emph {et~al.}(2017)\citenamefont {Silver}
  \emph {et~al.}}]{Silver2017Mastering}%
  \BibitemOpen
  \bibfield  {author} {\bibinfo {author} {\bibfnamefont {D.}~\bibnamefont
  {Silver}} \emph {et~al.},\ }\href {\doibase 10.1038/nature24270} {\bibfield
  {journal} {\bibinfo  {journal} {Nature}\ }\textbf {\bibinfo {volume} {550}},\
  \bibinfo {pages} {354} (\bibinfo {year} {2017})}\BibitemShut {NoStop}%
\bibitem [{\citenamefont {Sutton}\ \emph {et~al.}(1999)\citenamefont {Sutton}
  \emph {et~al.}}]{Sutton1999Policy}%
  \BibitemOpen
  \bibfield  {author} {\bibinfo {author} {\bibfnamefont {R.~S.}\ \bibnamefont
  {Sutton}} \emph {et~al.},\ }in\ \href@noop {} {\emph {\bibinfo {booktitle}
  {Proceedings of the 13th International Conference on Neural Information
  Processing Systems}}},\ \bibinfo {series and number} {NIPS'99}\ (\bibinfo
  {publisher} {MIT Press},\ \bibinfo {address} {Cambridge, MA, USA},\ \bibinfo
  {year} {1999})\ p.\ \bibinfo {pages} {1057–1063}\BibitemShut {NoStop}%
\bibitem [{\citenamefont {Schulman}\ \emph {et~al.}(2015)\citenamefont
  {Schulman} \emph {et~al.}}]{John2015Trust}%
  \BibitemOpen
  \bibfield  {author} {\bibinfo {author} {\bibfnamefont {J.}~\bibnamefont
  {Schulman}} \emph {et~al.}\ }(\bibinfo  {publisher} {JMLR.org},\ \bibinfo
  {year} {2015})\ p.\ \bibinfo {pages} {1889–1897}\BibitemShut {NoStop}%
\bibitem [{\citenamefont {Schulman}\ \emph {et~al.}(2017)\citenamefont
  {Schulman}, \citenamefont {Wolski}, \citenamefont {Dhariwal}, \citenamefont
  {Radford},\ and\ \citenamefont {Klimov}}]{schulman2017proximal}%
  \BibitemOpen
  \bibfield  {author} {\bibinfo {author} {\bibfnamefont {J.}~\bibnamefont
  {Schulman}}, \bibinfo {author} {\bibfnamefont {F.}~\bibnamefont {Wolski}},
  \bibinfo {author} {\bibfnamefont {P.}~\bibnamefont {Dhariwal}}, \bibinfo
  {author} {\bibfnamefont {A.}~\bibnamefont {Radford}}, \ and\ \bibinfo
  {author} {\bibfnamefont {O.}~\bibnamefont {Klimov}},\ }\href@noop {}
  {\bibfield  {journal} {\bibinfo  {journal} {arXiv preprint arXiv:1707.06347}\
  } (\bibinfo {year} {2017})}\BibitemShut {NoStop}%
\bibitem [{\citenamefont {Gidney}(2021)}]{stim_library}%
  \BibitemOpen
  \bibfield  {author} {\bibinfo {author} {\bibfnamefont {C.}~\bibnamefont
  {Gidney}},\ }\href@noop {} {\enquote {\bibinfo {title} {Stim: A fast and
  scalable quantum error correction simulator},}\ }\bibinfo {howpublished}
  {\url{https://github.com/quantumlib/stim}} (\bibinfo {year}
  {2021})\BibitemShut {NoStop}%
\bibitem [{\citenamefont {Morvan}\ \emph {et~al.}(2024)\citenamefont {Morvan}
  \emph {et~al.}}]{Morvan2024Phase}%
  \BibitemOpen
  \bibfield  {author} {\bibinfo {author} {\bibfnamefont {A.}~\bibnamefont
  {Morvan}} \emph {et~al.},\ }\href {\doibase 10.1038/s41586-024-07998-6}
  {\bibfield  {journal} {\bibinfo  {journal} {Nature}\ }\textbf {\bibinfo
  {volume} {634}},\ \bibinfo {pages} {328} (\bibinfo {year}
  {2024})}\BibitemShut {NoStop}%
\bibitem [{\citenamefont {Arute}\ \emph {et~al.}(2019)\citenamefont {Arute}
  \emph {et~al.}}]{Arute2019Quantum}%
  \BibitemOpen
  \bibfield  {author} {\bibinfo {author} {\bibfnamefont {F.}~\bibnamefont
  {Arute}} \emph {et~al.},\ }\href {\doibase 10.1038/s41586-019-1666-5}
  {\bibfield  {journal} {\bibinfo  {journal} {Nature}\ }\textbf {\bibinfo
  {volume} {574}},\ \bibinfo {pages} {505} (\bibinfo {year}
  {2019})}\BibitemShut {NoStop}%
\bibitem [{\citenamefont {Wu}\ \emph {et~al.}(2021)\citenamefont {Wu} \emph
  {et~al.}}]{Wu2021Strong}%
  \BibitemOpen
  \bibfield  {author} {\bibinfo {author} {\bibfnamefont {Y.}~\bibnamefont {Wu}}
  \emph {et~al.},\ }\href {\doibase 10.1103/PhysRevLett.127.180501} {\bibfield
  {journal} {\bibinfo  {journal} {Phys. Rev. Lett.}\ }\textbf {\bibinfo
  {volume} {127}},\ \bibinfo {pages} {180501} (\bibinfo {year}
  {2021})}\BibitemShut {NoStop}%
\bibitem [{\citenamefont {Gao}\ \emph {et~al.}(2025)\citenamefont {Gao} \emph
  {et~al.}}]{Gao2025Establishing}%
  \BibitemOpen
  \bibfield  {author} {\bibinfo {author} {\bibfnamefont {D.}~\bibnamefont
  {Gao}} \emph {et~al.},\ }\href {\doibase 10.1103/PhysRevLett.134.090601}
  {\bibfield  {journal} {\bibinfo  {journal} {Phys. Rev. Lett.}\ }\textbf
  {\bibinfo {volume} {134}},\ \bibinfo {pages} {090601} (\bibinfo {year}
  {2025})}\BibitemShut {NoStop}%
\end{thebibliography}%
\end{document}